\documentclass[prl,twocolumn,groupedaddress]{revtex4}

\usepackage{amsmath,bm}
\usepackage{graphicx}

\begin{document}

\title{Non-Fermi liquid behavior and superconductivity in the
dissipative model}


\author{P. C. Howell}
\author{A. J. Schofield}
\affiliation{Department of Physics and Astronomy, University of Birmingham, 
Birmingham B15 2TT, United Kingdom}

\date{\today}

\begin{abstract}
We study an exactly-solvable model which shows a zero-temperature
transition from a non-Fermi liquid to a Fermi liquid as a function of
particle density. The quantum critical point separating these two
states is not associated with the usual ordering transition of a
bosonic degree of freedom. We analyze the properties of this critical
point in terms of an effective density of states, and we use this
concept to generalize this transition to a class of related quantum
critical points. We study the superconducting instability of the
low-density (non-Fermi liquid) regime and we show that it has an
enhanced pair susceptibility relative to the Fermi liquid. We
comment on the applicability of our results to low-carrier density
superconductors.
\end{abstract}


\maketitle

In recent years there has been great interest in understanding the
physics of metallic states which are not Fermi liquids---motivated
largely by a host of experimental observations~\cite{santabarbara96}.  One
route to non-Fermi liquid behavior stems from proximity to a quantum
critical point~\cite{hertz76,millis93}, in other words a
zero-temperature phase transition which occurs as a parameter such as
pressure or chemical composition is varied. Examples include
CePd$_2$Si$_2$~\cite{julian96}
, and
CeCu$_{6-x}$Au$_x$~\cite{lohneysen94}
 which show quantum
critical behavior as pressure and gold concentration are varied
respectively. In both of these materials the critical point separates
a magnetically ordered state from a paramagnetic one.

The metallic state of the cuprates also shows non-Fermi liquid
properties. Furthermore there is evidence of a pseudogap energy scale
\cite{timusk99}, $T^*$, which is 
suppressed to zero temperature at some critical doping 
$p_\mathrm{crit}$~\cite{tallon01}, and it has been
suggested~\cite{tallon99} that this forms a quantum critical point
albeit masked by superconductivity. However, if this quantum critical
point is responsible for the non-Fermi liquid physics of the metallic
state, it differs from the above examples in that it does not appear
to have an associated ordered phase: thermodynamic 
measurements~\cite{loram94b} show that $T^*$ corresponds to a
crossover and not a phase transition.  It has been
argued~\cite{chakravarty00} that there may be a hidden order
parameter, but an alternative scenario is that the cuprates show
quantum criticality without associated order at finite temperature.
Such behavior is common in one dimension where thermal fluctuations
destroy finite-temperature order, but this motivates the search for
examples of zero-temperature phase transitions in $d>1$ without any
associated order. 

In this Letter we consider a model of dissipative fermions, first
studied by Wheatley~\cite{wheatley91}, which possesses a
zero-temperature transition but with no order other than the Fermi
surface itself. At low carrier density, this model is a non-Fermi
liquid metal but with a zero temperature transition to a Fermi liquid
at a critical density.  We characterize the model's behavior at the
critical point by the Sommerfeld coefficient and the paramagnetic
susceptibility as $T\rightarrow 0$.  We present an interpretation of
the phase diagram by introducing an effective density of states. This
allows us to consider more general metal to metal
transitions of this type. We also consider the superconducting
instability of this model. Although our starting model is artificial,
we speculate on the connection between models of this type and atom
traps containing neutral fermions as well as low carrier concentration
superconductors.

Our starting point is to consider a single particle of mass $M$
coupled to a bath of harmonic oscillators~\cite{caldeira81}. This bath 
has a spectral density $J(\omega)$ and was originally introduced to mimic
the dissipative effects of a real-world
environment~\cite{footnote1}. However, our philosophy here is to
introduce the oscillator bath to give the single particle a
non-trivial ({\it i.e.} non-local) dynamics in imaginary time while
retaining the exact solvability of the system.  This may be seen on
integrating out the oscillators, which leads to a single-particle
effective action $S_\mathrm{eff} = S_\mathrm{KE} +S_\mathrm{diss} $,
where $S_\mathrm{KE}=\frac{1}{2}\hbar M\beta \sum_n r_n
r_{-n}\,\omega_n^2$ is the kinetic energy and
\begin{equation}
S_\mathrm{diss} = \frac{2\hbar M\beta}{\pi} \!\sum_{n=-\infty}^\infty
\!\! r_n r_{-n}\!  \int_0^\infty \!{\rm d}\omega \,J(\omega)\,
\frac{\omega_n^2}{\omega(\omega^2+\omega_n^2)} .  
\end{equation} 
Here $r_n$ is the Fourier component of the path in imaginary time and
$\omega_n=\frac{2\pi\hbar\beta}{n}$ is a Matsubara frequency. One
usually chooses $J(\omega)=\eta\omega$ with a high-frequency cut-off
$\sim\!\omega_{\mathrm c}$, so that $S_\mathrm{diss}$ corresponds
classically to a frictional force proportional to velocity, $F=\eta
v$. In the first part of this paper we also make this choice, but
later we consider the consequences of more general spectral functions.

An exact expression can be found for the single-particle partition
function $z(\beta)$~\cite{wheatley91}. When the temperature is low
compared with the damping rate, $\tau_0^{-1}=2M/\eta$, so $\hbar
\beta/\tau_0 \gg 1$, this may be simplified to
\begin{equation}
z(\beta) = n_0 \,{\rm e}^{-\varepsilon_0 d\beta}
\left[1+\frac{d\pi\tau_0}{6\hbar\beta} + O({\beta^{-2}}) \right]
\label{partfunc} 
\end{equation} 
in $d$-dimensions, where
 $ n_0 = \left(M/2\pi\hbar\tau_0\right)^{\!d/2} $
has the dimensions of density
and $\varepsilon_0$ depends on the details of the cut-off in
$J(\omega)$ and will be absorbed as a shift in the band-edge. 
The dissipation defines an energy scale
$ \varepsilon_{\rm d} = {6 \hbar/d\pi \tau_0}$.

Having defined the properties of the single particle we now consider a
gas of such particles ({\it i.e.} fermions coupled to their own
independent oscillator baths). Antisymmetrizing appropriately for
fermions is most easily done using a path integral
approach~\cite{wheatley91} which gives the grand thermodynamic
potential for the dissipative gas 
\begin{equation} 
\Omega=\sum_{m=1}^\infty
(-1)^m \frac{{\rm e}^{m\beta\mu}}{m\beta} z(m\beta) \; . \label{omega}
\end{equation} 
Wheatley~\cite{wheatley91} solved this equation numerically to
determine the chemical 
potential $\mu$ and the susceptibility $\chi$, finding a $T=0$
phase transition from a low density ($n<n_0$) phase with Curie
susceptibility to a high density $(n>n_0$) phase with Pauli
susceptibility. Here we consider the system properties in the vicinity
of the critical point ($n\sim n_0$) and we find a generalization of
the model.

The chemical potential is given by the solution of 
\begin{equation} 
n=-\left(\frac{\partial\Omega}{\partial\mu}\right)_{\!T} =
\frac{n_0}{1+{\rm e}^{-\beta\mu}} + \frac{n_0}{\varepsilon_\mathrm{d} \beta} 
\ln(1+{\rm e}^{\beta\mu})  \; ;
\label{mueq} 
\end{equation} 
once we have determined $\mu$, the formulas for $\chi$ and $\gamma$
follow straightforwardly.  The solution of Eq.~\eqref{mueq} in the limit
$T\!\rightarrow\! 0$ depends on whether we approach the critical
density from higher or lower densities. 

Approaching the critical density from above 
$(\delta_n=\frac{n}{n_0}-1>0)$
we find
\begin{eqnarray}
\chi & = & \frac{n_0\mu_\mathrm{B}^2}{\varepsilon_\mathrm{d}}
\left(1+[\varepsilon_\mathrm{d} 
\beta-1]\,{\rm e}^{-\beta\varepsilon_\mathrm{d}\delta_n} \right) 
\label{chiabove} \;,  \\ 
\gamma & = & \frac{n_0k_\mathrm{B}^2}{\varepsilon_\mathrm{d}}
\left(\frac{\pi^2}{3} + 
(\beta \varepsilon_\mathrm{d})^3\delta_n^2{\rm e}^{-\beta\varepsilon_\mathrm{d} 
\delta_n}\right) 
\label{gammaabove} \; . 
\end{eqnarray} 
We see that proximity to the critical density sets a new scale
$\varepsilon_\mathrm{d} \delta_n$.  For low temperatures, $\beta 
\varepsilon_\mathrm{d} \delta_n
\gg 1$, the properties recover those of the free Fermi gas:
$\gamma$ and $\chi$ are constant. We call this the Fermi liquid regime,
although there are no Landau parameters since the particles do not
interact with each other. 

Approaching the critical density from below, $\delta_n <0$, we find
$\mu=k_{\mathrm B} T\ln(-\delta_n^{-1}-1)$ and hence for small
$\delta_n$
\begin{eqnarray}
\chi & = & - \frac{n_0\mu_\mathrm{B}^2}{\varepsilon_\mathrm{d}} 
\frac{\beta \delta_n}{1-\delta_n} \;, \label{chibelow} \\
\gamma & = & \frac{n_0k_\mathrm{B}^2}{\varepsilon_\mathrm{d}}
\left[\frac{\pi^2}{3}- \left(1-\frac{4}{\beta \varepsilon_\mathrm{d} \delta_n} 
\right) 
\ln^2 |\delta_n| \right]
\; . \label{gammabelow} \\
& & \nonumber
\end{eqnarray}
There is also a residual ground state entropy 
in the low density phase which is given by 
$S(T\!=\!0) = k_{\mathrm B} n_0
(|\delta_n|-|\delta_n|\ln|\delta_n|)$.

\begin{figure}
\includegraphics[width=86mm]{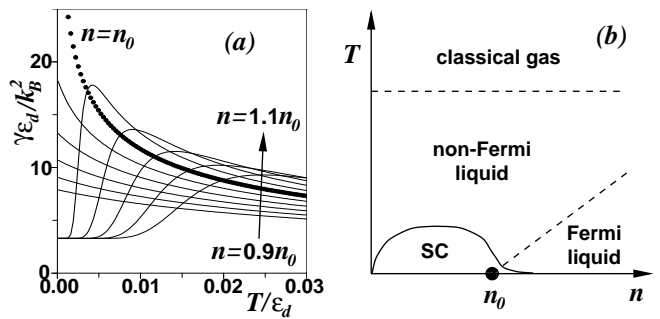}
{\caption{\small (a)  $\gamma(T)$ for a range of densities. (b) $T-n$
phase diagram 
of the dissipative gas. Solid lines are phase 
transitions, dashed lines are crossovers, SC = superconductor and the
circle is a quantum critical point.} 
\label{dissphase}}
\end{figure}

It is rather straightforward to see how this new phase of matter comes
about. Usually the transition to a quantum fluid occurs when the
thermal wavepacket of a particle starts to overlap with the
wavepacket of neighboring particles. In this model dissipation cuts
off the de Broglie wavelength such that for particle densities lower
than $n_0$ the quantum liquid regime is not entered at any
temperatures. For the case of a dissipative Bose gas one has the
possibility of a Bose metal phase~\cite{wheatley92}---the bosonic
analogue of the non-Fermi liquid state that results here for
dissipative fermions.

Having characterized the properties in the vicinity of the critical
density, we show in Fig.~\ref{dissphase}a the results of a numerical
calculation of the Sommerfeld coefficient as the critical density is
approached. This clearly shows the divergence in $\gamma$. It also
shows how the divergence is cut off at higher densities by the
crossover into the Fermi liquid regime. We have calculated the form of
the divergence at the critical point itself and find that $\chi\propto\ln
T$ and $\gamma\propto\ln^{2}T$. At a given $T$ the properties of the
high-density regime below a crossover scale $\delta_n\sim 
T/\varepsilon_\mathrm{d}$
qualitatively resemble those of the low-density regime, which we
illustrate by a dashed line on the phase diagram of
Fig.~\ref{dissphase}b \cite{footnote2}. There is a
second, high-temperature crossover which is set by the highest scale
in the oscillator bath $\beta \omega_c \sim 1$, above which 
no dissipation is felt by the particles.

Further insight into the nature of the phase diagram and the route to
generalizing this model can be found by considering the ``effective''
density of states. We use this term advisedly for the energy
eigenstates of a dissipative particle are not well defined---it
continuously interacts with the oscillator bath. However, since we
have a single-particle partition function, Eq.~\eqref{partfunc}, we can
use it to define an effective single-particle density of states
$g_{\rm eff}$ via $z(\beta)=\int_0^\infty {\rm d} E \,g_{\rm
eff}(E){\mathrm e}^{-\beta E}$. Thus the effective density of states
is formally the inverse Laplace transform of the single-particle
partition function. From here it is easy to show that the
thermodynamics can be obtained directly from $g_{\rm eff}$ by direct
substitution into the usual definition of the non-interacting
thermodynamic potential $\Omega = - T \int_0^\infty{\rm d} E \,g(E)
\ln\!\left[1+{\rm e}^{\beta(\mu-E)}\right]$.  The effective density of
states therefore
captures the thermodynamics of the dissipative gas, although it does
not contain information about correlations.

For the partition function of Eq.~\eqref{partfunc} we find an
effective density of states near the band-edge
\begin{equation} 
g_{\rm eff}(E+\varepsilon_0) =
\frac{n_0}{\varepsilon_\mathrm{d}} 
\left\{\delta\!\left(\frac{E}{\varepsilon_\mathrm{d}}\right)
+\Theta\!\left(\frac{E}{\varepsilon_\mathrm{d}}\right) \left[1+
O\!\left(\frac{E}{\varepsilon_\mathrm{d}}\right) \right]
\right\}  \, .
\end{equation} 
It is the initial delta function in the effective density of states
that leads to the non-Fermi liquid to Fermi liquid transition. The
delta function can accommodate $n_0$ fermions so for $n<n_0$ the Fermi
energy is pinned there and the entropy is clearly finite at $T=0$;
this is the non-Fermi liquid regime. When $n>n_0$ the Fermi energy
moves a distance $\delta_n \varepsilon_\mathrm{d}$ above the delta function, to
the region where $g_{\rm eff}$ is weakly energy-dependent, and the
system behaves like a degenerate Fermi gas. However, when
 $T\sim\delta_n\varepsilon_\mathrm{d}$ the Fermi function is
sufficiently broad 
that the occupancy of the delta function fluctuates and there is a
crossover to dissipative behavior.

We now consider extensions of this approach to consider a wider range
of Fermi liquid to non-Fermi liquid transitions that can be found for
dissipative fermions. The obvious extension is to go beyond Ohmic
dissipation and consider more general oscillator baths with power-law
spectral densities $J(\omega) = \eta \omega^\alpha$ up to a
high-frequency cut-off $\sim\!\omega_{\mathrm c}$, where $0<\alpha<2$.
Integrating out the oscillators gives
(say for $d=1$) the partition function
\begin{equation} 
z(\beta)=\left(\frac{M}{2\pi\beta}\right)^{\!\!1/2}
\,\prod_{n=1}^{\beta\omega_{\mathrm c}/2\pi}
\frac{\omega_n^2}{\omega_n^2+\omega_n^\alpha/\tau_0^{2-\alpha}}
\label{genprod} \; .
\end{equation} 
At low temperatures, $k_{\mathrm B}T\ll\hbar\tau_0^{-1}$, this is
approximately 
\begin{equation}
z(\beta)=n_0\left(\frac{\beta}{\tau_0}\right)^{\!\!\frac{1-\alpha}{2}}
{\rm e}^{-\varepsilon_0\beta}\,
\left[1+\frac{2-\alpha}{6}\frac{\pi\tau_0}{\hbar\beta} +O(\beta^{-2})
 \right] \; . 
\end{equation} 
The band edge shift $\varepsilon_0$ is again cut-off dependent. In terms
of the energy scale $\varepsilon_{\!\alpha}=6\hbar/(2-\alpha)\pi\tau_0$ the
inverse Laplace transform gives, to leading order in
$E/\varepsilon_{\!\alpha}$, 
\begin{equation} g_{\rm eff}(E+\varepsilon_0) = \left\{
\begin{array}{llr}
n_0 \big(\tfrac{E}{\varepsilon_{\!\alpha}}\big)^{\!\frac{\alpha-3}{2}}
& \mathrm{for} & \alpha > 1 \; ,\\ 
n_0\big[ \delta \big(\tfrac{E}{\varepsilon_{\!\alpha}}\big) +E\big] &
\mathrm{for} & \alpha=1 \; , \\
\!\begin{array}{l}\text{non-integrably} \\ 
\text{divergent} \end{array} & \mathrm{for} & \alpha<1 \; .
\end{array}
\right. 
\label{geffgen} 
\end{equation}
So, for $\alpha<1$ an infinite number of particles can be accommodated
in the ground state and the gas is a non-Fermi liquid for all particle
densities.  For $\alpha>1$ we always have conventional Fermi gas
behavior.  Thus only Ohmic dissipation gives rise to a quantum phase
transition between a Fermi liquid and a non-Fermi liquid.

Nevertheless, the notion of an effective density of states give us an
alternative route to explore other models. Since a quantum phase
transition will always be present when there is a 
delta function in $g_{\rm eff}$, we will work directly with a
generalized $g_{\rm eff}(E) =
n_0/\varepsilon_\mathrm{d} 
\left[\delta(E/\varepsilon_\mathrm{d})+(E/\varepsilon_\mathrm{d})^\nu
\,\Theta(E/\varepsilon_\mathrm{d}) \right] $, rather than explicitly
choose the spectral function of the oscillator bath. We must have
$\nu>-1$ for $g_{\rm eff}$ to be integrable. 
This gives the partition function 
\begin{equation}
z(\beta)=n_0\left[1+\Gamma(\nu+1)
\left(\varepsilon_\mathrm{d}T\right)^{\!\nu+1} \right] \; . 
\end{equation} 
Substituting this into Eq.~\eqref{omega} and using the properties of
the polylog Li$_\nu(x) = \sum_n\frac{x^n}{n^\nu}$
\cite{lewin81}, we find at $n=n_0$
\begin{eqnarray}
\chi & = & \frac{n_0\mu_\mathrm{B}^2}{\varepsilon_\mathrm{d}}
\frac{(\nu+1)^\nu}{\nu!} \,
\left(\varepsilon_\mathrm{d}\beta\right)^{-\nu}
\ln^{\nu+1}\varepsilon_\mathrm{d}\beta \label{chigen} \;, \\ 
\gamma & = &\frac{n_0k_\mathrm{B}^2}{\varepsilon_\mathrm{d}}
 \frac{(\nu+1)^{\nu+2}}{\nu!}
\,\left(\varepsilon_\mathrm{d}\beta\right)^{-\nu}  
\ln^{\nu+2}\varepsilon_\mathrm{d}\beta \; . \label{gammagen}
\end{eqnarray} 
Thus only for $\alpha\leq 0$ is the quantum critical
point characterized by divergences in $\chi$ and $\gamma$. Note that
for all $\alpha$ the Wilson ratio $\gamma/\chi$ is proportional to
$\ln T$.

We now address the
applicability of this model to real physical systems. Our method of
including all non-trivial effects via a bath of oscillators has
parallels with dynamical mean field theory where interactions
are included via a self-consistent frequency-dependent
self-energy~\cite{georges96,si99}. We have not sought a
self-consistent approach since our goal was to categorize all the
possible transitions in the model. Nevertheless our starting model may
be directly relevant to physical systems such as low density Fermi
gases which couple primarily to low energy bosonic modes rather than
each other. In particular, this necessitates short range
interactions. Examples could include neutral $^6$Li ions in atom
traps where the bosonic modes would be the ``optical molasses'' of
laser cooling giving a dissipative environment \cite{chu85}.

In a condensed matter context we apply this to very low carrier
density metals with strong electron-phonon coupling but in a
temperature window where the Coulomb repulsion does not dominate
(above any tendency to Wigner crystallize for example). One possible case
is SrTiO$_3$. This is a band insulator \cite{mattheis72}
but becomes a superconductor with $T_{\rm c}<1$~K
\cite{schooley65} on electron doping to a
concentration of about $10^{21}$~cm$^{-3}$.  The mechanism of
superconductivity in this material has long been in question not least
because superconductivity only appears {\em below} a critical
density. It is therefore natural to ask whether the non-Fermi liquid state
we have considered is unstable to pairing. 

We calculate the pair susceptibility in the dissipative model via the
pair propagator~\cite{schofield95}.  In $d$ dimensions the real-space
density matrix $\rho(r,\beta)$ for a dissipative particle is given
approximately by 
\begin{equation} 
\rho \simeq \left(\frac{m}{\pi\hbar}\right)^{\!d/2}
\!\left[\frac{1}{(\hbar\beta)^{d/2}}+\frac{1}{\tau_0^{d/2}}\right]
\mathrm{exp}\left[-\frac{r^2}{2}
\!\left(\frac{m}{\hbar^2\beta}+\frac{1}{\xi^2}\right)\right]
\label{denmatrix} 
\end{equation} 
where $\xi=\sqrt{\pi\hbar\tau_0/m\ln2}$ is a dissipation length. The
Matsubara Green's function can then be calculated using
$G_{\omega_n}= \int_0^\infty\rho(r,x)
\,{\rm e}^{(\mu+{\rm i}\hbar\omega_n)x}\,{\rm d} x$, which follows from the
definitions of $G_{\omega_n}$ and
$\rho$~\cite{schofield95}. This integral is only well defined for
$\mu<0$, which is the case for densities $n<\frac{n_0}{2}$, but it can
be  analytically continued to all $\mu$, as can be checked by
evaluating it for free fermions.  Substituting Eq.~\eqref{denmatrix}
with $d$=3 yields 
\begin{multline}
G_{\omega_n}(r) \simeq
-\left(\frac{m}{2\pi\hbar^2}\right)^{\!3/2}{\rm e}^{-r^2/2\xi^2}
\,{\rm e}^{-r\sqrt{\frac{2m}{\hbar^2}(-\mu-{\rm i}\hbar\omega_n)}} \\ \times
\left[\left(\frac{2\pi\hbar^2}{mr^2}\right)^{\!1/2} -
\left(\frac{\hbar}{\tau_0}\right)^{\!3/2}\frac{1}{\mu+{\rm i}\hbar\omega_n}
\right] \;.
\end{multline}

From the definition of the pair propagator
$K(r)=T\sum_{\omega_n}\!G_{\omega_n}(r)\,G_{-\omega_n}(r)$ we can
determine $T_{\mathrm c}$ in the usual way as the solution of
$1=\int{\rm d}\bm{r}\,  V(r) K(r,\beta)$. We consider a BCS-like
attractive potential with small finite width $r_0$. For $n<n_0$ this
leads to 
\begin{equation}
T_{\mathrm c} = V_\mathrm{eff}
\left(\frac{m}{\hbar\tau_0}\right)^{\!3/2}
\frac{1-\frac{2n}{n_0}}{\ln(\frac{n_0}{n}-1)}
\; , 
\end{equation} 
where the normalized potential $V_\mathrm{eff}$ depends on the
details of $V(r)$ at $r<r_0$. For $n>n_0$, on the other hand, $T_c$ is
heavily suppressed  and superconductivity does not occur for
arbitrarily small $V_\mathrm{eff}$. The dependence of $T_c$ on $n$ is
shown in Fig.~\ref{dissphase}b. The pairing enhancement can be traced
back to the delta function in $g_{\rm eff}(E)$. 

However, we can expect a different field dependence of the upper
critical field, $B_{\rm c2}(T)$, for superconductivity in the
dissipative model because of the appearance of the dissipation length
scale. A magnetic field
destroys superconductivity by dephasing the two electrons that make up
a Cooper pair; in a dissipative gas the electrons are already dephased
due to the oscillator bath and so we expect a field to have negligible
effect at small $T$. $B_\mathrm{c2}(T)$ is given by the solution
of $1=\int_{r>r_0}^\infty{\rm d} \bm{r} \,K(r,\beta) \,{\mathrm
e}^{-eB_{\mathrm \!c2}\,r^2/2\hbar}$ \cite{helfand66}. For $n<n_0$
we find just below $T_{\rm c}$ that $B_{\mathrm c2}=\frac{4\ln
2}{3\pi}\frac{m}{e\tau_0}\big(1-\frac{T}{T_{\rm c}}\big)$, and at low
temperatures $B_\mathrm{c2}(T) \sim \ln T$. In
practice this divergence in $B_\mathrm{c2}(T)$ would be cut off by Landau
quantization effects which we have not included.

So we expect a significant enhancement in the upper critical field in
the dissipative model compared with $B_\mathrm{c2}(0)/T_{\rm c}^2$ which
for a Fermi liquid is $2\pi^2\frac{\hbar}{e}
\big(\frac{m^*k_\mathrm{B}}{\hbar^2k_\mathrm{F}}\big)^2$.  In thin
films of SrTiO$_3$~\cite{leitner00} this ratio is enhanced by about an
order of magnitude relative to the Fermi liquid expression, where
$m^*=5.3m_0$ is determined from $\gamma$ \cite{ambler66}. Thus
SrTiO$_3$ qualitatively resembles the dissipative gas both in its
phase diagram and its relative insensitivity to applied magnetic fields.

To summarize, we have considered the Fermi liquid to non-Fermi liquid
zero-temperature transition in a model of dissipative fermions. This
critical point occurs without any associated ordering and may be
generalized via the concept of an effective density of states.  At low
temperatures the non-Fermi liquid shows an enhanced pairing
susceptibility and a phase diagram similar to the low density
superconductor SrTiO$_3$. More generally, this type of non-Fermi
liquid state could occur during laser cooling in fermionic atom traps.

We thank M. W. Long and I. V. Lerner for useful discussions. PCH is
supported by EPSRC and AJS by the Royal Society
and the Leverhulme Trust.


%
%



\end{document}